\documentclass[a4paper,twoside]{article}
\newcommand{\affil}[1]{$^{\rm #1}$}
\usepackage[authoryear]{natbib}
\bibpunct{(}{)}{;}{a}{}{,}
\usepackage{graphicx}
\usepackage{latexsym}
\date{} %Please leave the date blank

\newcommand{\mic}{\mbox{\,$\mu$m}}
\title{\large\bf\flushleft The Emergent Flux and Effective Temperature of $\delta$~Canis Majoris}
\author{\parbox{\textwidth}{\flushleft
\vspace{-0.5cm}
{\it J.~Davis\affil{A,D}, A.~J.~Booth\affil{B},
M.~J.~Ireland\affil{C}, A.~P.~Jacob\affil{A},
J.~R.~North\affil{A}, S.~M.~Owens\affil{A},
J.~G.~Robertson\affil{A}, W.~J.~Tango\affil{A} and P.~G.~Tuthill\affil{A}}\\
\vspace{0.4cm}
{\small \affil{A}\,School of Physics, University of Sydney, NSW 2006, Australia}\\
{\small \affil{B}\,Jet Propulsion laboratory, California Institute
of Technology, Pasadena, CA 91109, USA}\\
{\small \affil{C}\,Planetary Science, MS 150-21, Caltech, 1200 E. California Blvd, Pasadena, CA 91125, USA}\\
{\small \affil{D}\,Email: j.davis@physics.usyd.edu.au}}}
\vspace{0.4cm}
\begin{document}
\twocolumn[
\begin{changemargin}{.8cm}{.5cm}
\begin{minipage}{.9\textwidth}
\vspace{-1cm}
\maketitle
%
%
%%%%%%%%%%%%%     ABSTRACT    %%%%%%%%%%%%%
%Abstract of no more than 200 words here.
\small{\bf Abstract:} New angular diameter determinations for the
bright southern F8 supergiant $\delta$~CMa enable the bolometric
emergent flux and effective temperature of the star to be
determined with improved accuracy. The spectral flux distribution
and bolometric flux have been determined from published photometry
and spectrophotometry and combined with the angular diameter to
derive the bolometric emergent flux
$\mathcal{F}=(6.50\pm0.24)\times 10^{7}$\,Wm$^{-2}$ and the
effective temperature $T_{\mathrm{eff}}=5818\pm53$\,K. The new
value for the effective temperature is compared with previous
interferometric and infrared flux method determinations. The
accuracy of the effective temperature is now limited by the
uncertainty in the bolometric flux rather than by the uncertainty
in the angular diameter.

%%%%%%%%%%%%%     KEYWORDS    %%%%%%%%%%%%%
\medskip{\bf Keywords:} stars: atmospheres--stars: fundamental parameters--stars:
individual ($\delta$~CMa)--techniques: interferometric

%%%%%%%%DO NOT EDIT%%%%%%%%%%%%
\medskip
\medskip
\end{minipage}
\end{changemargin}
]
\small
%%%%%%%%EDIT FROM HERE%%%%%%%%%%%%

\section{Introduction}

In the determination of 32 stellar effective temperatures by
\citet{76code}, which is still the basis of the temperature scale
for hot stars, the coolest and faintest star, and the star with
the largest temperature uncertainty ($\pm$7\%) was the southern F8
supergiant $\delta$~CMa (HR2693, HD54605).  The effective
temperatures were determined by combining the angular diameters
measured with the Narrabri Stellar Intensity Interferometer (NSII)
\citep{74hbdanda} with flux distributions constructed from various
sources of calibrated photometry and spectrophotometry.  The
angular diameter of $\delta$~CMa determined with the NSII at a
wavelength of 443\,nm had an uncertainty of $\pm14$\% and this was
the dominant uncertainty in the effective temperature
determination. Because the angular diameter was the least
accurately determined with the NSII it has been a prime target for
the Sydney University Stellar Interferometer (SUSI)
\citep{99susi1} as a demonstration of the improvement achieved in
angular diameter measurements. The angular diameter has been
measured with SUSI at wavelengths of 442\,nm \citep{99susi2} and
700\,nm \citep{07rtp} with greatly improved accuracy.  In this
paper we use the angular diameter with revised fluxes obtained
from published photometry and spectrophotometry to determine the
bolometric emergent flux and the effective temperature for
$\delta$~CMa with significantly improved accuracy.  The accuracy
is now limited by the uncertainty in the determination of the
bolometric flux received from the star after correction for
interstellar extinction.  The new directly determined temperature
is also compared with the effective temperature determined by the
infra-red flux method (IRFM).

The emergent flux at the surface of a star per unit wavelength
interval ($\mathcal{F}_{\lambda}$) is given by

\begin{equation}
\mathcal{F}_{\lambda} =
\frac{4}{\theta_{\mathrm{LD}}^{2}}f_{\lambda}
\end{equation}
where $\theta_{\mathrm{LD}}$ is the true limb-darkened angular
diameter of the star and $f_{\lambda}$ is the flux per unit
wavelength interval received at the Earth from the star at
wavelength $\lambda$, corrected for atmospheric and interstellar
extinction. The effective temperature of the star
($T_{\mathrm{eff}}$) is then given by

\begin{equation}
\sigma T_{\mathrm{eff}}^{4} = \mathcal{F} =
\int_{0}^{\infty}\mathcal{F}_{\lambda}d\lambda =
\frac{4}{\theta_{\mathrm{LD}}^{2}}\int_{0}^{\infty}f_{\lambda}d\lambda
\label{eq:efftemp}
\end{equation}
where $\sigma$ is the Stefan-Boltzmann radiation constant and
$\mathcal{F}$ is the bolometric emergent flux at the stellar
surface.

Thus, a knowledge of the limb-darkened angular diameter of the
star, and the flux distribution received from it, leads to a
direct determination of $T_{\mathrm{eff}}$. $f_{\lambda}$ can be
obtained from flux-calibrated photometry and spectrophotometry,
and $\theta_{\mathrm{LD}}$ can be obtained by interferometric
measurements. In the following sections we will consider the
determination of these two quantities for $\delta$ CMa, and
finally their combination to give $\mathcal{F}$ and
$T_{\mathrm{eff}}$.

\section{The Angular Diameter}\label{sec:angdiam}

\begin{table*}[t]
\begin{center}
  \caption{The uniform-disk angular diameter of $\delta$~CMa determined with the NSII and with
  SUSI.  $\lambda$ is the effective wavelength and $\Delta\lambda$ the spectral bandwidth of the
  measurement.    $V^{2}_{0}$ is the extrapolated value of visibility squared at zero baseline from the
  uniform-disk angular diameter fit to the observed values of $V^{2}$.  $\sigma$\% is the percentage
  uncertainty in the uniform-disk angular diameter.  The SUSI values for 442\,nm are revised
  values from a re-processing of the original data (see text).}
  \label{tab:udads}
  \begin{tabular}{cccccc}
\hline Instrument & $\lambda$ & $\Delta\lambda$ & $V^{2}_{0}$ & $\theta_{\mathrm{UD}}$ &  $\sigma$\% \\
 & (nm) & (nm) & & (mas) &  \\
 \hline
 NSII & 443.0 & 8 & $0.93\pm0.18$ & $3.29\pm0.46$   & 14.0 \\
 SUSI & 442.0 & 4 & $0.917\pm0.024$ & $3.41\pm0.10$ &  2.6 \\
 SUSI & 442.0 & 4 & $0.880\pm0.031$ & $3.37\pm0.15$ &  4.5 \\
 SUSI & 696.6 & 80 & $1.003\pm0.012$ & $3.457\pm0.024$ &  0.7 \\
\hline
\end{tabular}
\end{center}
\end{table*}

The values for the equivalent uniform-disk angular diameter,
$\theta_{\mathrm{UD}}$, determined with the NSII and with SUSI
have been discussed by \citet{07rtp}.  The values for
$\theta_{\mathrm{UD}}$ determined with the NSII and at 695.6\,nm
with SUSI, taken from their Table~3, are listed in
Table~\ref{tab:udads}.  The SUSI values in Table~\ref{tab:udads}
for 442\,nm are the revised values discussed by \citet{07rtp}
obtained after re-processing the observational data with the
omission of observations not bracketed by a calibrator.  All the
determinations involved a two-parameter fit to the measured values
of the square of the fringe visibility ($V^{2}$), the fitting
parameters being the equivalent uniform-disk angular diameter and
the value of $V^{2}$ at zero baseline, $V_{0}^{2}$. The values of
$V_{0}^{2}$ for each of the fits are included in
Table~\ref{tab:udads}.

\citet{07rtp} determined, for each value of the uniform-disk
angular diameter, the true, limb-darkened angular diameter of
$\delta$~CMa by applying the appropriate correction factor
interpolated from the tabulation of \citet{00dtb}.  The effective
temperature was initially taken to be $6000\pm200$\,K, based on a
number of values in the literature, with $\log{g}=0.6$ and
[Fe/H]$= 0.19$ from \citet{85landl}, for the interpolation. After
the effective temperature had been determined to be 5818\,K,
following the procedure discussed in Section~\ref{sec:efet}, the
limb-darkening correction factors were checked using the revised
temperature with the same values for $\log{g}$ and [Fe/H].  The
only change was for 695.6\,nm with an increase from 1.050 to
1.051.  Although this has negligible effect, reducing the
temperature by only 2\,K, the revised value has been used in the
final analysis.  The correction factors and the resulting values
for the limb-darkened angular diameter, based on the values for
the uniform-disk angular diameter in Table~\ref{tab:udads}, are
listed in Table~\ref{tab:ldads}.

\begin{table}[h]
\begin{center}
  \caption{The limb-darkened angular diameter of $\delta$~CMa.  $\rho_{\lambda}$ is the
  ratio of limb-darkened to uniform-disk angular diameter used to convert the uniform-disk
  angular diameters in Table~\ref{tab:udads} to the limb-darkened angular diameters
  in this table (details are given in the text).}
  \label{tab:ldads}
  \begin{tabular}{cccc}
\hline Instrument & $\lambda$ & $\rho_{\lambda}$ & $\theta_{\mathrm{LD}}$ \\
 & (nm) &  & (mas) \\
 \hline
 NSII & 443.0 &  1.099 & $3.62\pm0.51$ \\
 SUSI & 442.0 &  1.100 & $3.75\pm0.11$ \\
 SUSI & 442.0 &  1.100 & $3.70\pm0.17$ \\
 SUSI & 695.6 &  1.051 & $3.633\pm0.026$ \\
\hline
\end{tabular}
\end{center}
\end{table}

As noted by \citet{07rtp} the uncertainty in the NSII value for
the limb-darkened angular diameter is large and covers all three
values determined with SUSI.  The new 695.6\,nm value differs from
the two 442\,nm values but, although the two 442\,nm values agree
with one another, we believe that the new value is the most
reliable.  The reasons for this belief have been discussed in
detail by \citet{07rtp}.  In brief, the 442\,nm observations were
made during the commissioning phase of SUSI and significant
improvements had been made in the observing, calibration and
seeing correction techniques prior to the 695.6\,nm observations.
In particular, not all the 442\,nm observations of $\delta$~CMa
were bracketed by a calibrator and, as reported by \citet{07rtp},
a re-analysis omitting these data has led to the revised values
for the uniform-disk angular diameters listed in
Table~\ref{tab:udads} and, consequently, to the revised values for
the limb-darkened angular diameters listed in
Table~\ref{tab:ldads}. The revised values lie within
$\sim1.1\sigma$ and $\sim0.4\sigma$ of the 695.6\,nm result.

In Table~\ref{tab:udads} the extrapolated values of $V^{2}$ at
zero baseline, $V^{2}_{0}$, for the uniform-disk angular diameter
fits to the observational data are listed.  The values for the
NSII 443\,nm and SUSI 442\,nm observations are all less than the
expected value of unity for a single star.  The NSII value is
consistent with the value for a single star because of its large
uncertainty. However, the two SUSI values at 442\,nm are
significantly less than unity leading to speculation
\citep{99susi2} that $\delta$~CMa might be a binary system with a
significantly fainter companion.  As noted by \citet{99susi2} the
fact that the observational points are a reasonable fit to the
curve for a single star at 442\,nm suggests that, if the star is a
binary system, the $V^{2}$ values at each baseline are averaged
over a range in position angles \citep{67hbdar}.  The $V^{2}_{0}$
values are consistent with a companion 3.25 magnitudes fainter
than $\delta$~CMa at 442\,nm. SUSI data at both 442\,nm and
695.6\,nm have been examined for potential position angle
variations that would confirm the presence of a companion with a
negative result. The value of $V^{2}_{0}$ of 1.003$\pm$0.012 at
695.6\,nm is consistent with $\delta$~CMa being a single star. The
possibility of a faint hot companion significantly affecting the
blue measurements while having a negligible effect on the red
measurements has been considered. While such a scenario would
result in a larger magnitude difference at 700\,nm than at 442\,nm
the maximum effect would be a difference of 5 magnitudes resulting
in a value for $V^{2}_{0}$ of 0.98.  This differs by $\sim$2
standard deviations from the observed value.

After careful examination of the data and reduction procedures we
have concluded that the 695.6\,nm result is correct and that there
is no observable companion.  The results from the 442\,nm
observations must now be regarded as suspect due to the
difficulties of calibration and correction of the larger seeing
effects at the shorter wavelength.  The original agreement between
the two 442\,nm results, while encouraging at the time, is thought
to be fortuitous.  This is supported by the fact that the omission
of data not bracketed by a calibrator has resulted in significant
changes to the 442\,nm uniform-disk angular diameters and brought
them closer to the 695.6\,nm result.  The longer wavelength result
also has a smaller correction for limb darkening and is therefore
less model dependent.  For the determination of the bolometric
emergent flux and the effective temperature of $\delta$~CMa we
adopt the angular diameter result for 695.6\,nm.

\section{The Integrated Flux}

The integrated flux for $\delta$~CMa has been determined following
the procedure used by \citet{76code} but with a revised estimate
for interstellar extinction, improved flux calibrations, and some
more recent visual and infrared data.  Following \citet{76code} it
is appropriate to divide the flux measurements into three
wavelength regions: ultraviolet, visible and infrared since they
rely on different techniques for their calibration. The boundary
between the visual and infrared regions has been moved from
810\,nm to 860\,nm due to the availability of new extended visual
data and the three regions are discussed individually in
Sections~\ref{sec:uv} to \ref{sec:ir}.

Since $\delta$~CMa is reddened by interstellar extinction,
corrections must be applied in order to determine the emergent
flux and effective temperature.  This is discussed in the
following section.

\subsection{Correction for Interstellar
Extinction}\label{sec:extinc}

The observed value of (\textit{B}-\textit{V}) for $\delta$~CMa is
$+0.67$ \citep{66hljetal} and the intrinsic value for an F8~Iab
star is $+0.56$ \citep{82s-k} giving a colour excess of
\textit{E}(\textit{B}-\textit{V}) equal to 0.11. This is close to
the value of 0.12 used by \citet{76code} which was based on an
intrinsic value of $+0.55$ by \citet{66hlj} but, as pointed out by
\citet{82fernie}, reddenings determined in this way are unreliable
because the reddening line so nearly parallels the
(\textit{U}-\textit{B})$_{0}$ v. (\textit{B}-\textit{V})$_{0}$
intrinsic sequence for supergiants.  In fact, $\delta$~CMa lies
almost on the intrinsic sequence but closer to G0 than F8.

\citet{67feinstein} has studied the young southern cluster
Collinder~121 and, from ten early-type main-sequence stars,
deduced that \textit{E}(\textit{B}-\textit{V}) for the cluster
does not exceed 0.03.  He also associated $\delta$~CMa with the
cluster.  However, more recent studies \citep{00kat, 07bur} place
Collinder 121 at a distance greater than 1000\,pc with a
foreground moving association of stars at a distance of
$\sim700$\,pc.  With the Hipparcos parallax giving its distance as
$550\pm170$\,pc it is likely that $\delta$~CMa is a member of this
latter group with \textit{E}(\textit{B}-\textit{V}) of the order
of 0.03.  Using spectrum synthesis and model atmospheres
\citet{75pandb} have also derived a value of 0.03 for
\textit{E}(\textit{B}-\textit{V}) for $\delta$~CMa.
\citet{72schmidt} derived a value of 0.05 and \citet{91mcw} used
\textit{A$_{V}$} = 0.10, equivalent to
\textit{E}(\textit{B}-\textit{V})$\sim$0.03, derived from `forcing
consistency between all de-reddened colors'.

These alternative approaches to the evaluation of
\textit{E}(\textit{B}-\textit{V}) point to a value of
\textit{E}(\textit{B}-\textit{V}) of 0.03 and it is clear that the
value adopted by \citet{76code} is incorrect.  We adopt
\textit{E}(\textit{B}-\textit{V}) = 0.030 with an uncertainty of
\textbf{$\sim\pm0.015$}.

The interstellar extinction curve used by \citet{76code}, and
listed in their Table~2, has been adopted to correct the UV
fluxes.  For the visual and near infrared fluxes
($\lambda\lambda$0.33--1.0\,$\mu$m) the average interstellar
extinction curve given by \citet{82s-k} has been used.  For
wavelengths in the range $\lambda\lambda$1.0--13.0\,$\mu$m the
interstellar extinction law given by \citet{85randl} has been
adopted.  Beyond 13.0\,$\mu$m interstellar extinction is
negligible for $\delta$~CMa.

\subsection{The Ultraviolet Flux}\label{sec:uv}

The flux below 330\,nm makes only a small contribution to the
total flux ($<1.3$\%).  We have therefore adopted the flux
reported by \citet{76code} obtained with the OAO-2 satellite.
Application of the revised reddening correction and its
uncertainty, as discussed in Section~\ref{sec:extinc}, gives the
flux for the wavelength interval 0--330\,nm equal to
$(0.063\pm0.013)\times10^{-9}$\,Wm$^{-2}$.

\subsection{The Visual Flux}\label{sec:vis}

\citet{76code} based the visual flux for the wavelength interval
330--810\,nm on the relative spectrophotometric measurements of
\citet{74dandw}.  Subsequently \citet{87kiehling} published
spectrophotometry for $\delta$~CMa for the wavelength range
325--865\,nm.  The observations were made at equidistant intervals
of 1\,nm with a resolution of 1\,nm.  The published spectral
energy distributions are averaged over band-passes 5\,nm wide and
are tabulated every 5\,nm.  The \citet{74dandw} data were
published for 25 selected 5\,nm pass-bands in the wavelength range
330--808\,nm.  In this section we compare these two sets of data
and the empirical MILES fluxes of \citet{06sanchez}.

\begin{figure}[t]
\begin{center}
\includegraphics[scale=0.54]{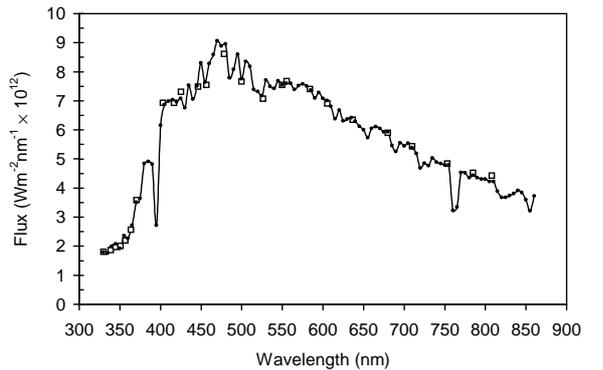}
\caption{The dereddened visual flux distribution for $\delta$~CMa
for the wavelength range 330-860\,nm.  Key: $\bullet$~(joined by
smoothed line) \citet{87kiehling}; $\Box$~\citet{74dandw}. Details
are given in the text.} \label{fig:visflux}
\end{center}
\end{figure}

\citet{76code} used the spectrophotometric calibration of
$\alpha$~Lyr (Vega) by \citet{70oke} to convert the relative
spectrophotometry of \citet{74dandw} into a relative absolute flux
distribution.  Here the more recent spectrophotometric calibration
of Vega by \citet{85hayes} has been used.  Following
\citet{76code} the resulting relative absolute flux distribution
has been scaled by the flux ratio corresponding to the
monochromatic magnitude of $\delta$~CMa relative to Vega at
550\,nm (1.779) measured by Davis (private communication). It has
then been converted to fluxes using the value for the flux from
Vega at 550.0\,nm of $3.56\times10^{-11}$\,Wm$^{-2}$nm$^{-1}$
\citep{95meg}.  The published \citet{87kiehling} spectrophotometry
is already in the form of a relative absolute flux distribution
based on the Hayes calibration.  It has been scaled and flux
calibrated in exactly the same way as the Davis \& Webb relative
absolute flux distribution.  The two sets of calibrated flux
distributions are in excellent agreement with an RMS difference
computed from the wavelengths in common of $<1.1$\% with no
systematic differences over the wavelength range in common
(330--808\,nm).  The two flux distributions are shown in
Figure~\ref{fig:visflux}.

The flux distribution in the MILES library of empirical spectra
\citep{06sanchez} for $\delta$~CMa has also been considered.
Unfortunately the listed MILES fluxes have been de-reddened on the
assumption of \textit{E}(\textit{B}-\textit{V}) = 0.209.  The
fluxes have been corrected to those for the value of
\textit{E}(\textit{B}-\textit{V}) = 0.03 and calibrated to give
absolute fluxes in the same way as the Davis \& Webb and Kiehling
data.  The wavelength cover of the MILES flux distribution is
355-740\,nm, less than the Kiehling range of 325-865\,nm.  It is
tabulated at 0.9\,nm intervals with a resolution of 0.23\,nm
compared with the data by Kiehling, which were averaged over 5\,nm
intervals and tabulated every 5\,nm.  To compare the two
distributions the fluxes were integrated for the common wavelength
range of 355-740\,nm. The integrated fluxes agree to within 1\%.
The difference is small compared with the uncertainty, which is
dominated by the uncertainty in the reddening.  The Kiehling flux
distribution covers a greater wavelength range and extends to the
ultraviolet data at the short wavelength end and, for these
reasons, it has been used to determine the integrated visual flux.

The visual flux integration has been extended to 860\,nm, since
Kiehling has data points to 865\,nm, rather than terminate it at
810\,nm like \citet{76code}. Application of the interstellar
extinction (reddening) correction to the individual flux values
and integration of the resulting dereddened flux distribution,
gives the total flux for the wavelength interval 330--860\,nm
equal to $(3.05\pm0.13)\times10^{-9}$Wm$^{-2}$.  The quoted
uncertainty is solely due to the uncertainty assigned to
\textit{E}(\textit{B}-\textit{V}) but uncertainties in the
relative absolute flux distribution of $\delta$~CMa, in the
monochromatic magnitude used for scaling, and in the absolute flux
calibration must also be considered.

The agreement between all three flux distributions considered, and
the good agreement between the integrated fluxes for the
wavelength range 355-740\,nm for the MILES and Kiehling flux
distributions, suggests that the uncertainty in the relative
absolute flux distributions is at the 1\% level. The uncertainty
in the monochromatic magnitude difference used for scaling the
$\delta$~CMa flux distribution is estimated to be $\sim$1\% and
\citet{95meg} claims $\pm$0.7\% for the flux calibration at
550\,nm.  The largest uncertainty by far is $\pm4.3$\% due to the
uncertainty in \textit{E}(\textit{B}-\textit{V}). The
uncertainties are independent and have been combined accordingly
to give a resultant uncertainty of $\pm$4.6\%.  The estimated
total flux for the wavelength interval 330--860\,nm is
$(3.05\pm0.14)\times10^{-9}$Wm$^{-2}$.

\subsection{The Infrared Flux}\label{sec:ir}

More extensive IR data exist than were available to
\citet{76code}, and these have been used to improve the value for
the integrated flux in this region.  Estimating the total IR flux
involves considering data from a number of sources in different
forms and with differing calibrations.  For this reason the
contributions for the wavelength intervals 0.86--1.0\,$\mu$m,
1.0--2.5\,$\mu$m and 2.5-22.5\,$\mu$m have been considered
separately and the results summed. The contribution for
wavelengths longer than 22.5\,$\mu$m is negligible ($<$0.01\% of
the total flux).

\begin{figure}[h]
\begin{center}
\includegraphics[scale=0.45]{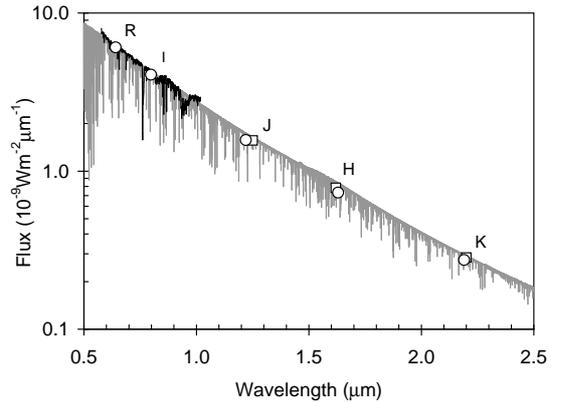}
\caption{The flux distribution for $\delta$~CMa for the wavelength
range 0.5--2.5\,$\mu$m.  Key: Black line - \citet{94danks} with
slope correction; $\Box$ - \textit{RIJHK} broad-band photometry
using absolute flux calibration by \citet{98bessell}; $\circ$ -
\textit{JHK} broad-band photometry using absolute flux calibration
of \citet{95meg}; Gray line - fitted model atmosphere.  Details
are given in the text.} \label{fig:irflux01}
\end{center}
\end{figure}

\citet{94danks} give relative spectrophotometry for $\delta$~CMa
for the wavelength range 0.58--1.02\mic.  We have calibrated their
data by comparing it with the 7 pass bands of \citet{74dandw} in
the overlap region and with the \citet{87kiehling}
spectrophotometry from 0.58--0.865\mic. The distributions are in
agreement at 0.62\mic\,\,and the calibration results in a
wavelength dependence of $\sim6.1$\% per 100\,nm in the overlap
region 0.58--0.86\mic.  This slope correction has been applied to
the Danks \& Dennefeld data for the wavelength range
0.58--1.0\mic.  The revised distribution shows good agreement with
the \textit{R} and \textit{I} broad-band fluxes discussed below
and this can be seen in Figure~\ref{fig:irflux01}. Corrections for
reddening have been applied to the resulting flux distribution for
the wavelength range 0.86--1.0\mic\,\,and the flux integrated. The
uncertainty in the integrated flux due to the uncertainty in
\textit{E}(\textit{B}-\textit{V}) is less than for the
330--860\,nm wavelength range but the uncertainty in the
calibration of the fluxes is larger.  The overall uncertainty is
estimated to be $\pm$3.6\%.  The resulting estimate for the total
flux in the wavelength interval 0.86-1.0\mic\,\,is
(4.34$\pm$0.16)$\times$10$^{-10}$\,Wm$^{-2}$.

In the 1-2.5\mic\,\,interval the only data available are
broad-band \textit{JHK} photometric measurements.  Although
broad-band IR photometry is not ideally suitable for accurate flux
determinations, since it is strongly affected by atmospheric
extinction which changes the effective spectral pass bands in ways
that are difficult to take into account \citep{96bliek}, we have
shown that flux-calibrated \textit{RIJHK} photometry is consistent
with the slope-corrected Danks and Dennefeld flux distribution.

In view of the sparsity of observational data in the
1-2.5\mic\,\,interval a model atmosphere flux distribution has
been fitted to the dereddened data and used to derive integrated
fluxes for this spectral range. The flux in the wavelength range
1.0--2.5\mic\,\,has been represented by fitting the flux
distribution for a NextGen Model \citep{99ng2} to the
slope-corrected Danks \& Dennefeld flux distribution plus
flux-calibrated \textit{R}, \textit{I}, \textit{J}, \textit{H} and
\textit{K} broad-band photometry between 0.7\mic\,\,and 2.2\mic.
The fitted model has a temperature of 5800\,K ($\log{g} = 1.0$,
[Fe/H] = 0) which is essentially the same as the effective
temperature of 5818\,K determined in this work for the star.  The
photometric magnitudes have been selected and flux calibrated as
follows.  For \textit{R} and \textit{I} the magnitudes by
\citet{80cousins} have been adopted as they are more reliable than
those by \citet{66hljetal} for these bands \citep{07bessell}. They
have been calibrated using the absolute flux calibration of
\citet{98bessell}.  For \textit{J}, \textit{H} and \textit{K} the
magnitudes were adopted from examination of the photometry of
\citet{66hljetal} (\textit{J} and \textit{K}), \citet{74glass}
(\textit{J}, \textit{H} and \textit{K}), \citet{81engels}
(\textit{J}, \textit{H} and \textit{K}) and \citet{90carter}
(\textit{J}, \textit{H} and \textit{K}). The \textit{JHK}
photometry was flux calibrated using the absolute flux
calibrations of both \citet{95meg} and \citet{98bessell}. The
model flux distribution was fitted by eye to the observational
data by means of a scaling factor and the fitted curve and data
points are shown in Figure~\ref{fig:irflux01}.  The uncertainty in
the integrated flux for the range 1-2.5\mic\,\,is based on the
combination of the estimated uncertainty in the model fit
($\pm2.5$\%), the uncertainty in the absolute flux calibration
(taken as $\pm2$\% as given for \textit{JHK} by \citet{95meg}),
and the uncertainty in the dereddening correction ($\pm1.3$\%).
The integrated flux for the wavelength range 1.0--2.5\mic\,\,is
$(1.32\pm0.05)\times10^{-9}$Wm$^{-2}$.

For the spectrum longward of 2.5\mic\,\,the \textit{L} and
\textit{M} photometric bands lie in the 2.5--5.0$\mu$m range and
there are IRAS Point Source fluxes at 12, 25, 60 and
100\mic\,\,\citep{88iras} and IRAS Low-Resolution Spectra (LRS)
covering $\sim7.7-22.7$\mic\,\,\citep{86iras}.  All these data lie
significantly above the fluxes for the model atmosphere fitted to
the 0.7-2.2$\mu$m interval.  Since it is unclear whether the
observed flux is from the star or surrounding material we have
evaluated the flux longward of 2.5\mic\,\,in two ways.

Firstly, we have integrated the fitted model fluxes from
2.5-22.5\mic.  The upper wavelength limit corresponds to the long
wavelength end of the IRAS LRS spectra. The integrated flux for
the 2.5-22.5\mic\,\,range is 0.164$\times10^{-9}$Wm$^{-2}$.

\begin{figure}[h]
\begin{center}
\includegraphics[scale=0.52]{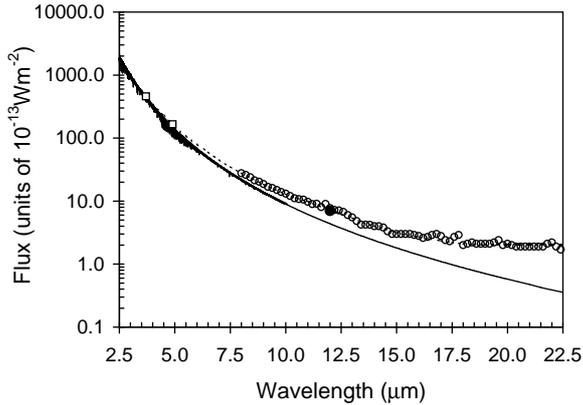}
\caption{The dereddened flux distribution for $\delta$~CMa for the
wavelength range 2.5--22.5\,$\mu$m.  Key: Solid line - NextGen
5800\,K model; Dashed line - Smooth curve fitted to observational
data; $\Box$ - broad-band \textit{L} and \textit{M} fluxes;
$\circ$ - IRAS LRS fluxes; $\bullet$ - IRAS Point Source flux.
Details are given in the text.} \label{fig:irflux02}
\end{center}
\end{figure}

The second approach has been to use the broad-band \textit{L} and
\textit{M} fluxes, the IRAS Point Source flux at 12\mic, with the
IRAS LRS fluxes and to bridge the gaps in the data by drawing a
smooth curve through them.  The observational data have been
assembled as follows.  The magnitudes for the \textit{L} and
\textit{M} photometric bands have been adopted from examination of
the photometry of \citet{66hljetal} (\textit{L}), \citet{74glass}
(\textit{L}), \citet{81engels} (\textit{L} and \textit{M}) and
\citet{90carter} (\textit{L}). The magnitudes were flux calibrated
using the calibration of \citet{95meg} for \textit{L} and
\citet{66hlj} for \textit{M} and corrected for reddening.  The
IRAS Point Source flux at 12\mic\,\, was reduced by 4.1\% as
proposed by \citet{96cohen} to bring it into line with their
absolute calibration.   The IRAS LRS fluxes have been corrected
using the factors determined by \citet{92cww} and are claimed to
be accurate to better than 2\% \citep{04price}.  The dereddened
and flux calibrated data were plotted against wavelength and a
smooth curve drawn through them.  The curve was then tabulated at
regular intervals across the wavelength range 2.5-22.5\mic.  The
\textit{L} flux lies $\sim8$\% above the model curve, the
\textit{M} flux $\sim14$\% above and the IRAS LRS flux at
8\mic\,\,$\sim28$\% above.  Figure~\ref{fig:irflux02} shows the
measured flux data, the curve for the model atmosphere flux
distribution that was fitted to the wavelength interval
0.7-2.2\mic, and the smooth curve drawn through the data.  The
integrated flux in the interval 2.5-22.5\mic\,\, for the curve
drawn through the data is 0.194$\times10^{-9}$Wm$^{-2}$.

The difference in the integrated flux between the two approaches
is $\sim0.03\times10^{-9}$Wm$^{-2}$.  We have adopted the mean
value ($0.18\times10^{-9}$Wm$^{-2}$) with an uncertainty of
$\pm0.02\times10^{-9}$Wm$^{-2}$.  This corresponds to an
uncertainty of $\pm0.4$\% in the total flux received from the star
and translates to an uncertainty of $\pm0.1$\% in the effective
temperature ($\sim\pm6$\,K).

The total IR flux for the wavelength range 0.86--22.5\,$\mu$m is
the sum of the integrated fluxes for the intervals 0.86--1.0\mic,
1.0--2.5\mic, 2.5--22.5\mic.  As noted earlier the flux for
wavelengths longer than 22.5\,$\mu$m is negligible. The
uncertainty in the total flux was estimated by simply summing the
individual errors since they are likely to be dominated by
calibration uncertainties and are thus systematic and not
independent.  The resultant value for the total IR flux is
(1.93$\pm$0.09)$\times$10$^{-9}$\,Wm$^{-2}$.

\subsection{The Total Flux}

\begin{figure}[h]
  \begin{center}
  \includegraphics[scale=0.48, angle=0]{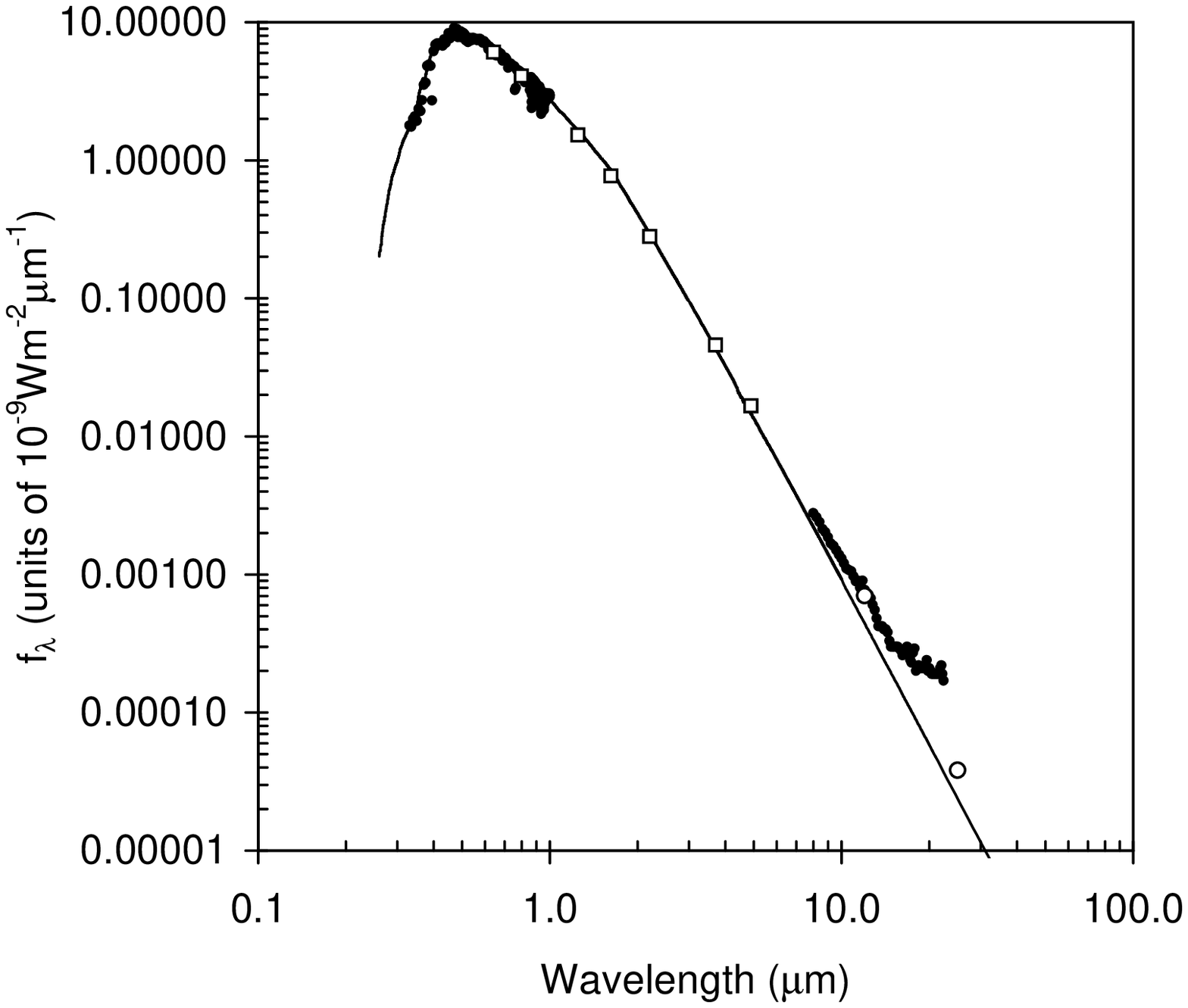}
   \caption{The interstellar extinction-corrected visual and IR flux for
   $\delta$ CMa.  The data are represented in three wavelength ranges by
   dots: 0.33--0.86\,$\mu$m \citep{87kiehling}, 0.86--1.0\,$\mu$m \citep{94danks}
   and 8.0--22.5\,$\mu$m IRAS LRS \citep{86iras}.  $\Box$ - fluxes from broad-band
   photometry; $\circ$ - IRAS Point Source fluxes.  The line represents the fluxes averaged
   over 5\,nm bands for the NextGen 5800\,K model atmosphere fitted to the observational
   data in the 0.7-2.2\,$\mu$m interval.  Details of the calibration and integration
   of the fluxes are given in the text.}
  \label{fig:flux}
  \end{center}
\end{figure}

Figure~\ref{fig:flux} shows the overall extinction-corrected flux
distribution, from the visible to the IR, made up of the data used
in integrating the visual and IR contributions.

The total flux received from $\delta$~CMa after correction for
interstellar extinction is given by summing the contributions from
the ultraviolet, visual and infrared regions of the spectrum.  The
contributions are listed  with the total flux of
(5.04$\pm$0.17)$\times10^{-9}$\,Wm$^{-2}$ in Table~\ref{tab:flux}.
In assessing the uncertainty in the total flux, the uncertainties
from the separate wavelength regions have been considered to be
independent.

\begin{table}[h]
\begin{center}
\caption{The extinction corrected integrated fluxes for
$\delta$~CMa in each spectral band plus the total integrated flux
from the star.}
\begin{tabular}{c c}
\hline
Wavelength & Flux \\
(nm) & ($10^{-9}$\,Wm$^{-2}$) \\
\hline
0--330 & 0.06 $\pm$ 0.01 \\
330--860 & 3.05 $\pm$ 0.14 \\
860--$\infty$ & 1.93 $\pm$ 0.09\\
& \\
Total flux & 5.04 $\pm$ 0.17 \\
\hline
\end{tabular}
\label{tab:flux}
\end{center}
\end{table}

The total integrated flux is significantly less than the value of
(6.01$\pm$0.27)$\times10^{-9}$\,Wm$^{-2}$ derived by
\citet{76code}. This is attributable to the revised value for
\textit{E}(\textit{B}-\textit{V}).  The uncertainty has been
reduced due to additional flux measurements in the visual and
infrared and improvements in the absolute flux calibration.

\section{The Emergent Flux and Effective
Temperature}\label{sec:efet}

The bolometric emergent flux and effective temperature for
$\delta$~CMa are found by substituting the limb-darkened angular
diameter and the extinction-corrected total flux received from the
star in equation~\ref{eq:efftemp}.  The bolometric emergent flux
$\mathcal{F}$ is (6.50$\pm$0.24)$\times$10$^{7}$\,Wm$^{-2}$ and
the effective temperature $T_{\mathrm{eff}}$ is 5818$\pm$53\,K.
The dominant source of uncertainty in $T_{\mathrm{eff}}$ is from
the integrated flux (0.8\%), with a smaller contribution from the
angular diameter (0.4\%).

\section{The Radius and Luminosity}\label{sec:randl}

The angular diameter can be combined with the parallax of the star
to determine the stellar radius, and the combination of radius and
bolometric emergent flux gives the stellar luminosity.
Unfortunately the Hipparcos parallax for $\delta$~CMa is of low
accuracy with $\pi=1.82\pm0.56$\,mas.  Nevertheless a value for
the radius has been calculated and is listed in
Table~\ref{tab:props} together with the bolometric emergent flux
and effective temperature.  The large fractional uncertainty of
$\sim\pm31$\% in the parallax dominates the fractional uncertainty
in the radius.  The luminosity depends on the square of the radius
so the percentage error is doubled and the luminosity, with an
uncertainty of $\sim\pm62$\%, is of little value and has not been
listed in Table~\ref{tab:props}.

\begin{table}
\begin{center}
\caption{The physical parameters determined for $\delta$~CMa.}
\begin{tabular}{lc}
\hline
\multicolumn{1}{c}{Parameter} & Value \\
\hline
Bolometric emergent & \\
flux ($\mathcal{F}$) ($10^{7}$\,Wm$^{-2}$) & $6.50\pm0.24$ \\
Effective temperature ($T_{\mathrm{eff}}$) (K) & $5818\pm53$ \\
Radius ($R/R_{\odot}$) & $215\pm66$  \\
\hline
\end{tabular}
\label{tab:props}
\end{center}
\end{table}

\section{Discussion}
Previous determinations of the angular diameter, bolometric flux
and effective temperature for $\delta$~CMa are listed in
Table~\ref{tab:temps} with the new values presented in this paper.
The only other direct determination of effective temperature for
this star is by \citet{76code} who obtained a value of
$T_{\mathrm{eff}}=6110\pm430$\,K using the angular diameter
determined with the NSII.   The higher value for the effective
temperature determined by Code et al. is almost entirely due to
the value of \textit{E}(\textit{B}-\textit{V}) they adopted. While
Code et al. underestimated the uncertainty in the bolometric flux
by not including an allowance for the uncertainty in the
interstellar extinction corrections, the uncertainty in their
temperature is primarily due to the uncertainty in the NSII
angular diameter. The NSII angular diameter contributed $\pm$6.9\%
to the uncertainty in the effective temperature compared with
$\pm$1.1\% from the bolometric flux they derived.

The new temperature determination presented here lies within the
uncertainty of the Code et al. value but has a substantially
reduced uncertainty.  The bolometric flux is now the dominant
source of uncertainty, primarily due to the interstellar
extinction uncertainty in the visible and dependence on broad-band
photometry in the near IR.

\begin{table}
\begin{center}
\caption{The limb-darkened angular diameter
$\theta_{\mathrm{LD}}$, interstellar extinction corrected
bolometric flux $f$, and effective temperature $T_{\mathrm{eff}}$
for $\delta$~CMa from various sources.  Key: 1 -- \citet{76code};
2 -- \citet{80black}; 3 -- \citet{91mcw}); 4 -- This work.}
\begin{tabular}{clll}
\hline
Key & \multicolumn{1}{c}{$\theta_{\mathrm{LD}}$}& \multicolumn{1}{c}{$f$} & \multicolumn{1}{c}{$T_{\mathrm{eff}}$}  \\
   & \multicolumn{1}{c}{\scriptsize{(mas)}} & \multicolumn{1}{c}{\scriptsize{(10$^{-9}$Wm$^{-2}$)}} & \multicolumn{1}{c}{\scriptsize{(K)}}  \\
\hline
1 & $3.60 \pm 0.50$ & 6.01$\pm$0.27 & $6110 \pm 430$  \\
2 & 3.56 & 6.0 & 6143  \\
3 & & 5.14 & 5855  \\
4 & $3.633 \pm 0.026$ & 5.04$\pm$0.17 & $5818 \pm 53$  \\
\hline
\end{tabular}
\label{tab:temps}
\end{center}
\end{table}

The infra-red flux method (IRFM) \citep{77bands} has been used to
determine an angular diameter and effective temperature for
$\delta$~CMa by \citet{80black}.  Using a value of
$(6.0\pm0.3)\times10^{-9}$\,Wm$^{-2}$ for the bolometric flux from
the star \citep{77bands}, they derive an angular diameter of
3.56\,mas and an effective temperature of 6143\,K.  No uncertainty
is quoted for the angular diameter, and the effective temperature
is suggested to be accurate to about 2\%.  Their angular diameter
only differs from the measured value presented here by $\sim$2.1\%
which would only affect the temperature by $\sim$1.1\%.  The
difference in temperature of $\sim$5.5\% is mainly due to the
larger bolometric flux which is essentially the same as that
derived by \citet{76code}.  As discussed in
Section~\ref{sec:extinc} it is believed that the corrections
applied by \citet{76code} were too large although it is not known
what corrections were applied by \citet{80black}.

\citet{91mcw} has also determined an effective temperature for
$\delta$~CMa using the IRFM.  Using a similar value for the
interstellar extinction as in this paper, he derived a value for
the bolometric flux of 5.14$\times10^{-9}$\,Wm$^{-2}$ and an
effective temperature of 5855\,K.  Although no uncertainties were
quoted, the value for the effective temperature is a weighted mean
of four values, ranging from 5776\,K to 5953\,K, each determined
from a different IR pass-band. Both the bolometric flux and
effective temperature lie within the uncertainties of the new
values presented here.

\section{Conclusion}

We have determined new and improved values for the bolometric
emergent flux ((6.50$\pm$0.24)$\times10^{7}$\,Wm$^{-2}$) and
effective temperature (5818$\pm$53\,K) for the F8 supergiant
$\delta$~CMa using a new interferometric angular diameter measured
with SUSI. The uncertainty in the effective temperature has been
reduced from $\pm$7.0\% to $\pm$0.9\%. It has been shown that
precise temperatures can be obtained by combining angular
diameters measured interferometrically with bolometric flux
distributions assembled from a wide range of sources, with the
dominant uncertainty now coming from the bolometric flux
determination.

\section*{Acknowledgments}
This work has been carried out as part of the SUSI programme which
has been funded jointly by the Australian Research Council and the
University of Sydney, with additional support from the Pollock
Memorial Fund and the Science Foundation for Physics within the
University of Sydney. It has made use of the SIMBAD database,
operated by CDS, Strasbourg, France and VizieR \citep{vizier}. The
authors are grateful to the referee for a number of suggestions
which have improved both the paper and the accuracy of the
bolometric flux determination.  MJI acknowledges the support of an
Australian Postgraduate Award, JRN and APJ the support of
University Postgraduate Awards, and APJ and SMO the support of
Denison Postgraduate Awards.

\end{document}